\documentclass[aps,twocolumn,superscriptaddress,numerical]{revtex4-2}
\usepackage{graphicx}
\usepackage{amsmath}
\usepackage{amsbsy}
\usepackage{array}
\usepackage{amssymb}
\usepackage{verbatim}
\usepackage{caption} 
\usepackage{subcaption} 
\usepackage{hyperref}
\usepackage{epstopdf}
\usepackage{comment}

\usepackage{natbib}
\hypersetup{
	colorlinks=true,  
	linkcolor=blue,   
	citecolor=blue,   
	urlcolor=blue,    
	pdfborder={0 0 0}, 
}

\begin{document}
\title{A General Physics-Informed Machine Learning Method for Determining the Steerability of Arbitrary Entangled States}

\author{Yanning Jia}
\affiliation{School of Mathematical Sciences, Beijing University of Posts and Telecommunications, Beijing 100876, China}
\affiliation{Key Laboratory of Mathematics and Information Networks, Beijing University of Posts and Telecommunications, Ministry of Education, China}
\affiliation{State Key Laboratory of Networking and Switching Technology, Beijing University of Posts and Telecommunications, Beijing, 100876, China}

\author{Fenzhuo Guo}\email{gfenzhuo@bupt.edu.cn}
\affiliation{School of Mathematical Sciences, Beijing University of Posts and Telecommunications, Beijing 100876, China}
\affiliation{Key Laboratory of Mathematics and Information Networks, Beijing University of Posts and Telecommunications, Ministry of Education, China}
\affiliation{State Key Laboratory of Networking and Switching Technology, Beijing University of Posts and Telecommunications, Beijing, 100876, China}

\author{Mengyan Li}
\affiliation{School of Mathematical Sciences, Beijing University of Posts and Telecommunications, Beijing 100876, China}
\affiliation{Key Laboratory of Mathematics and Information Networks, Beijing University of Posts and Telecommunications, Ministry of Education, China}
\affiliation{State Key Laboratory of Networking and Switching Technology, Beijing University of Posts and Telecommunications, Beijing, 100876, China}

\author{Haifeng Dong}
\affiliation{School of Instrumentation Science and Opto-Electronics Engineering, Beihang University, Beijing,100191,China}

\author{Fei Gao}
\affiliation{State Key Laboratory of Networking and Switching Technology, Beijing University of Posts and Telecommunications, Beijing, 100876, China}

\begin{abstract}
Determining whether an arbitrary entangled state is steerable is a challenging task, as it requires assessing the existence of a local hidden-state (LHS) model across all relevant measurement settings. In this work, we develop a Physics-Informed Machine Learning method that reformulates the construction of the LHS model as a differentiable parameter optimization problem. By leveraging the vmap vectorized transformation in JAX for efficient batch-wise sampling, we efficiently explore the continuous measurement space. Furthermore, physical constraints are embedded into the parameterization, and automatic differentiation is used for gradient-based optimization to obtain the optimal LHS model. We validate our method on two-qubit Werner and two-qutrit isotropic states. For Werner states, our approach marks the first numerical success in precisely reproducing analytical visibility bounds under three Pauli measurements, arbitrary projective measurements (PVMs), and arbitrary positive operator-valued measurements (POVMs). For isotropic states, we achieve the known analytical bounds under arbitrary PVMs and explore the steerability under arbitrary POVMs, suggesting that POVMs can offer an advantage over PVMs in revealing the steerability of such states.
\end{abstract}

\maketitle

\section{Introduction}
Quantum steering is an important form of nonclassical correlation arising from local measurements on entangled states, with significant applications in a variety of information-processing tasks \cite{PhysRevLett.98.230501,PhysRevLett.113.140501, PhysRevApplied.23.014057,PhysRevA.87.012336, Chang_2024, Máttar_2017}. Originally introduced by Schrödinger \cite{Schrodinger_1935}, the concept of steering was later rigorously formalized by Wiseman et al \cite{PhysRevLett.98.140402}. When two parties share an entangled quantum state, the system is said to exhibit quantum steering if the post-measurement assemblages obtained by a trusted party, conditioned on local measurements performed by an untrusted party on the other subsystem, cannot be described by any local hidden-state (LHS) model. Since not all entangled states are steerable \cite{Saunders2010}, how can one determine whether a given entangled state is steerable? More precisely, an entangled state is steerable if there exists at least one measurement setting that generates an assemblage incompatible with any LHS model. Consequently, determining steerability necessitates a search over the entire measurement space to rule out the existence of a universal LHS model.


Implementing such a search is highly challenging. The continuous nature of the measurement space, together with the difficulty of constructing a universal response function for the LHS model across different measurements, makes exhaustive verification impractical. For certain classes of entangled states, analytical approaches have been successfully developed to determine their steerability. Notable examples include symmetric states \cite{PhysRevLett.98.140402} and arbitrary two-qubit states \cite{PhysRevA.93.022121} under projective measurements (PVMs), as well as Bell-diagonal and Werner states under general positive operator-valued measurements (POVMs) \cite{PhysRevA.101.042125, PhysRevLett.132.250201}. For more general states, numerical methods have been introduced to explore their steerability. Semidefinite programming (SDP) was the first to provide a systematic numerical framework for searching LHS models under arbitrary measurements \cite{PhysRevLett.117.190401, PhysRevLett.117.190402}. However, SDP is inherently limited to finite measurement samples and cannot generate universal response functions across the full measurement space. Recently, machine learning has emerged as a powerful tool in quantum information science \cite{PhysRevLett.120.240402, h7qp-63dg, PhysRevA.107.062409, Krivachy2020}. As shown in Ref. \cite{PhysRevA.111.052446}, neural networks were employed to determine quantum steerability. These approaches face the same limitations as SDP. Consequently, existing numerical methods cannot saturate the known analytical bounds when determining the steerability of two-qubit Werner states.

To address the aforementioned limitations, in this work we present a general Physics-Informed Machine Learning (PIML) method to determine the steerability of an arbitrarily given entangled state, building on the local hidden-variable (LHV) model construction framework of Ref.~\cite{PRXQuantum.6.020317} and extending it to the LHS setting. In the present LHS setting, the construction of the LHS model is reformulated as a differentiable parameter optimization problem. Our method implements a dynamic “sampling–evaluation–update” iterative loop, as introduced in Ref.~\cite{PRXQuantum.6.020317}. First, we utilize the vmap technique in JAX to achieve high-throughput, real-time random sampling. In each iteration, new measurement directions are dynamically drawn from the continuous measurement space, ensuring that the parameter search gradually traverses the entire space and effectively approximates the requirement of arbitrary measurements. Second, physical constraints are directly embedded into the parameterized architecture. The response function $p(a|x, \lambda)$ is represented as an expansion over orthonormal basis functions across the measurement space, ensuring generalization to unseen measurement directions while adhering to physical logic. Additionally, the hidden states $\sigma(\lambda)$ are parameterized via Cholesky decomposition, ensuring that every iteration strictly satisfies the physical requirements of density matrices, including positivity and unit trace. With these physical components, the corresponding LHS model is constructed, and the loss function is defined as the average trace distance between the LHS assemblages and the quantum assemblages. Leveraging automatic differentiation, exact gradients of the loss function with respect to the model parameters are obtained, enabling gradient descent to update the parameters in a self-consistent manner. Through this iterative cycle, an optimal LHS model is ultimately obtained. When the loss function converges to zero, it indicates that the model successfully reproduces the quantum assemblages, thereby certifying the entangled state as unsteerable at the computational level.

To evaluate the effectiveness of our method, we apply it to analyze the steerability of two-qubit Werner states and two-qutrit isotropic states. For these states, steerability depends on the noise level, and the critical visibility defines the threshold above which the state remains steerable. For two-qubit Werner states, our approach is the first numerical method to precisely recover the analytical visibility bounds reported for three Pauli measurements \cite{PhysRevA.80.032112}, arbitrary PVMs \cite{PhysRevA.93.022121}, and arbitrary POVMs \cite{PhysRevLett.132.250201}. For two-qutrit isotropic states, our numerical results achieve the analytical visibility bound under arbitrary PVMs \cite{PhysRevLett.98.140402}. While analytical bounds for general POVMs are lacking, our numerical method provides a clear indication that the critical visibility under POVMs is lower than that under PVMs, suggesting that POVMs may offer an advantage over PVMs in revealing the steerability of these states.

\section{Preliminaries}
First, we provide a more detailed description of quantum steering and the LHS model in the simplest bipartite scenario. We consider a one-sided device-independent scenario: Alice's side is the untrusted party, while Bob's side is the trusted party.

\begin{figure}
\centering
\begin{subfigure}[b]{0.45\linewidth}
    \centering  
    \includegraphics[width=\linewidth]{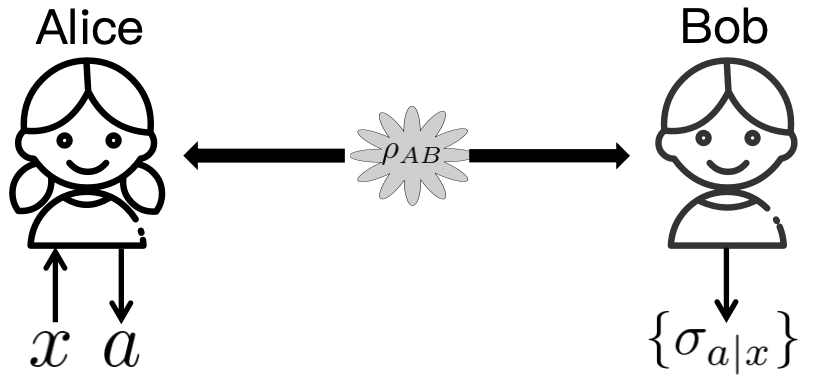}
    \caption{}  
    \label{fig2a}
\end{subfigure}
\hfill
\begin{subfigure}[b]{0.45\linewidth}
    \centering  
    \includegraphics[width=\linewidth]{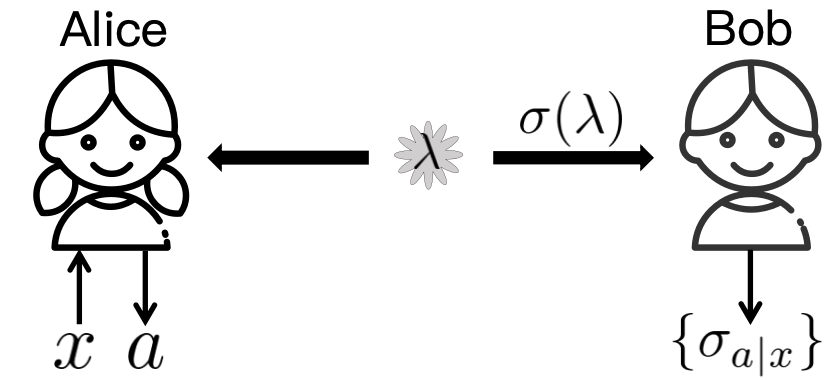}
    \caption{}  
    \label{fig2b}
\end{subfigure}
\caption{Bipartite Scenarios. (a) The Steering Scenario: Alice and Bob share an unknown entangled state $\rho_{AB}$. Alice's device acts as an untrusted black box, where the measurements $M_{a|x}$ she performs are completely unknown; Bob's side is trusted. We can determine whether Alice can steer Bob's state by obtaining the assemblage $\{\sigma_{a|x}\}$ from Bob's side. (b) The LHS Model: This model describes a situation where a source sends a hidden variable $\lambda$ to Alice and a corresponding hidden state $\sigma(\lambda)$ to Bob, thereby generating the assemblage $\{\sigma_{a|x}\}$.}
\label{fig2}
\end{figure}

In this setup, Alice and Bob share an unknown quantum state $\rho_{AB}$. Alice performs a local measurement labeled by index $x$ chosen from some family $\{M_{a|x}\}_{a,x}$, with $M_{a|x} \geq 0$ and $\sum_a M_{a|x} = I$. It should be noted that the measurements $M_{a|x}$ performed by Alice are completely unknown to us. The possible post-measurement states for Bob's system are given by the unnormalized state assemblage $\{\sigma_{a|x}\}$ (see Fig. \ref{fig2}(a)). According to quantum theory, the elements of this assemblage can be obtained through the following expression:

\begin{equation}
\sigma^{QM}_{a|x} = \text{tr}_A\left[(M_{a|x} \otimes I) \rho_{AB}\right].
\label{1}
\end{equation}

A steering scenario can be fully characterized by this assemblage. The probability $p(a|x)$ that Alice obtains outcome $a$ given measurement $x$ icalculated as $p(a|x) = \text{tr}(\sigma^{QM}_{a|x})$.

An LHS model refers to the situation where a source sends hidden variable $\lambda$ to Alice and the corresponding hidden state $\sigma_\lambda$ to Bob (see Fig. \ref{fig2}(b)). The unnormalized post-measurement state assemblage on Bob's side is composed of the elements:
\begin{equation}
\sigma^{LHS}_{a|x} = \int d\lambda\mu(\lambda) \, p(a|x,\lambda) \, \sigma_\lambda,
\label{2}
\end{equation}
where $\mu(\lambda)$ is a probability density function over hidden variable $\lambda$ shared between Alice and Bob, and $p(a|x,\lambda)$ is a response function for Alice.

An given assemblage is said to demonstrate steering if it does not admit a decomposition of the form Eq. \eqref{2}. Otherwise, the assemblage is unsteerable. An entangled state is steerable if and only if the assemblages obtained by arbitrary local measurements cannot be reproduced by any LHS model.

\section{Physically Constrained Parameterization of the LHS Model}
As we have introduced the LHS model in the previous section, constructing such a model requires determining three key components: the hidden-variable $\lambda$, the response function $p(a|x,\lambda)$, and the hidden state $\sigma(\lambda)$. In the following, we will describe how to construct these elements, which will allow us to effectively optimize the model in the subsequent steps. Since the measurement space is continuous and infinite, it is impractical to assign a distinct hidden variable $\lambda$ to every possible measurement setting. Instead, we adopt the construction introduced in Ref.~\cite{PRXQuantum.6.020317}, employing a finite number of hidden variables $\{\lambda_i\}_{i=1}^{N_{hidden}}$ and a universal response function that can generate outcome probabilities for arbitrary measurements.

\subsection{Parameterization of Hidden States with Physical Constraints}

For the hidden states $\sigma(\lambda)$, each hidden variable $\lambda$ corresponds to a density matrix defined in the Hilbert space of Bob’s subsystem. To ensure physical validity, each $\sigma(\lambda)$ must satisfy the positivity ($\sigma(\lambda) \ge 0$) and unit-trace ($\text{Tr}[\sigma(\lambda)] = 1$) requirements. 

In practice, directly enforcing these constraints during optimization is inefficient. To address this, our PIML method adopts a reparameterization scheme by introducing an unconstrained complex matrix $M_\lambda$ and defining:
\begin{equation}
\sigma_\lambda = \frac{M_\lambda M_\lambda^\dagger}{\mathrm{Tr}[M_\lambda M_\lambda^\dagger]}.
\end{equation}

By utilizing this Cholesky-like decomposition, the positivity and normalization of $\sigma_\lambda$ are strictly enforced by the model's architecture itself throughout the gradient descent process.

\subsection{Vectorization of the Measurement Space}

A crucial step in our PIML method is to map arbitrary quantum measurements into a vectorized coordinate system that the differentiable model can process. We consider a general qudit POVM $M_x$ consisting of operators $\{M_{a|x}\}_a$ satisfying $M_{a|x} \geq 0$ and $\sum_a M_{a|x} = I$. Using the convention employed in Ref.~\cite{PRXQuantum.6.020317}, these Hermitian operators can be expanded in terms of the generalized traceless Gell-Mann matrices $G_\mu$ and the identity matrix $I$:
\begin{equation}
M_{a|x} = M_{a|x}(\vec{g}) \equiv \frac{1}{\sqrt{2d}} \sum_{\mu=0}^{d^2-1} g_{\mu} G_{\mu}.
\end{equation}
Here, we set $G_0 = \sqrt{\frac{2}{d}} I$ such that:
\begin{equation}
\text{tr}(G_{\mu} G_{\nu}) = 2 \delta_{\mu\nu}, \quad \text{for all } \mu, \nu \in \{0,1,\ldots,d^2-1\}.
\end{equation}

Through this expansion, each measurement operator $M_{a|x}$ is uniquely represented by a Gell-Mann vector $\vec{g}^a \in \mathbb{R}^{d^2}$. 

Since any POVM can be expressed as a convex combination of extremal POVMs, we focus on extremal POVMs which contain at most $d^2$ elements \cite{DAriano_2005}. Thus, an extremal POVM $M_x$ is fully parameterized by at most $(d^2 - 1) \times d^2$ parameters, enabling the PIML method to traverse the continuous measurement space.

\subsection{Response Function via Orthonormal Basis Expansion}

To satisfy the probability axioms ($0 \le p(a|x,\lambda) \le 1$ and $\sum_a p(a|x,\lambda) = 1$) within a differentiable framework, the response probabilities $p(a|x,\lambda)$ are obtained by applying a \texttt{softmax} transformation to the functions $f_a(\vec{g}^a,\lambda)$, as in Ref.~\cite{PRXQuantum.6.020317}:
\begin{equation}
p(a|x,\lambda) = \frac{e^{f_a(\vec{g}^a,\lambda)}}{\sum_{a' = 1}^{O} e^{f_{a'}(\vec{g}^{a'},\lambda)}},
\end{equation}
where $O$ is the number of outcomes.

It is important to emphasize that a POVM $M_x = \{M_{a|x}\}_{a=1}^{O}$ is not characterized by a single vector, but rather by a collection of Gell-Mann vectors:
\[
\mathcal{G}_x = (\vec{g}^1, \vec{g}^2, \dots, \vec{g}^O),
\]
where each $\vec{g}^a \in \mathbb{R}^{d^2}$ represents the generalized Bloch vector of the corresponding measurement operator. Therefore, although the response function is written as $f_a(\vec{g}^a,\lambda)$, the basis functions used in its expansion are defined over the entire POVM space $\mathcal{G}_x$, allowing the model to capture correlations among different measurement outcomes induced by the completeness constraint.

To ensure the model generalizes to unseen measurement directions in a physically consistent manner, the measurement rule $f_a(\vec{g}^a,\lambda)$ is expanded, as in Ref.~\cite{PRXQuantum.6.020317}, using a set of orthonormal basis functions $\{B_m(\mathcal{G}_x)\}_{m=1}^N$:
\begin{equation}
f_a(\vec{g}^a,\lambda) = \sum_{m=1}^{N} c_a^m(\lambda)\, B_m(\mathcal{G}_x).
\end{equation}

The coefficients $c_a^m(\lambda)$ are determined by the hidden variable $\lambda$, which forms an $M \times N$ matrix:
\[
\mathcal{\lambda} =
\begin{bmatrix}
c_1^1(\lambda) & \cdots & c_1^N(\lambda) \\
\vdots & \ddots & \vdots \\
c_{M}^1(\lambda) & \cdots & c_{M}^N(\lambda)
\end{bmatrix}.
\]
This basis expansion allows the PIML model to approximate arbitrary response functions efficiently.

 \subsection{Examples of Response Function Construction}

For clarity, we present two representative examples: qubit PVMs ($d=2$), for which we follow the construction in Ref.~\cite{PRXQuantum.6.020317}, and qutrit POVMs ($d=3$). These examples demonstrate how measurement operators can be mapped into real vector spaces and how response functions are constructed over these representations.

\subsubsection{response function for qubit PVMs}
In the case of qubits ($d=2$), the projective measurement \(M_x\) is dichotomic with \(M_x = \{M_{0|x},M_{1|x}\}\). Each measurement operator \(M_{a|x}\) can be linearly expressed in terms of Pauli matrices $\{\sigma_x,\sigma_y,\sigma_z\}$ and the identity matrix $I$ as:
\[
M_{a|x} = \frac{I + (-1)^a \left(x \cdot \sigma_x + y \cdot \sigma_y + z \cdot \sigma_z\right)}{2}.
\]

Thus, each measurement operator \(M_{a|x}\) can be represented by a Bloch vector $\vec{g}^a = (x, y, z)$. The response function \(p(a|x,\lambda)\) then becomes a function constructed from the Bloch vector:
\begin{align}
&p(0|x,\lambda) = softmax[f_0(\vec{g}^0,\lambda)], \\  \notag
&p(1|x,\lambda) = softmax[f_1(\vec{g}^1,\lambda)], \\ \notag
&p(0|x,\lambda) + p(1|x,\lambda) = 1.
\end{align}

Due to the probability normalization condition, it suffices to determine only the measurement rule $f_0(\vec{g}^0)$.

Given the properties of dichotomic projective measurements, each measurement operator corresponds to a Bloch vector, and these vectors satisfy the relation $\vec{g}^1 = -\vec{g}^0$. Considering that deterministic measurement rules ($p(a|\cdot) \in \{0,1\}$) already encompass all possible scenarios, we have $p(1|x,\lambda) = 1 - p(0|x,\lambda)$ on one hand; on the other, the inherent constraints of deterministic rules further restrict the functional form. This leads to the conclusion that the measurement rule function $f$ satisfies $f(-\vec{g}^a,\lambda) = 1 - f(\vec{g}^a,\lambda)$. Based on this, we expand $f_0(\vec{g}^0)$ in the $\vec{g}^0$ space using odd spherical harmonics. This approach not only satisfies the odd symmetry requirement but also leverages orthogonality for efficient approximation. By introducing a order $D$, we denote the vector composed of odd spherical harmonics up to order $D$ as $\vec{B}^D(\vec{g}^0)$ (the specific form for $D=3$ is given below):
\begin{align}
    \vec{B}^3(\vec{g}^0) \sim (x, y, z, xyz, \dots, 3x^2y - y^3),
    \\ \notag
    \vec{g}^0 = (x, y, z).
\end{align}

When considering the measurement rule for a single outcome, the hidden variable reduces to a vector $\vec{\lambda} \in \mathbb{R}^N$ (instead of a matrix), whose dimension is given by:
\[
N = \frac{1}{2}(D+1)(D+2).
\]
To satisfy the requirement of non-vanishing gradients in numerical optimization, we employ a sigmoid function to transform the mapping into probabilistic form:
\begin{align}
&p(0|x,\lambda) = \mathrm{sigmoid}\left[\sum_{m=1}^{N}\vec{B}_m^D(\vec{g}^0) \cdot \vec{\lambda}\right],
\\ \notag
&\mathrm{sigmoid}[y] = \frac{1}{1+e^{-y}}.
\end{align}

\subsubsection{response function for qutrit POVMs}

In the qutrit systems ($d=3$), we consider extremal POVMs, which contain up to $d^2 = 9$ outcomes, i.e.,
$M_x = \{ M_{a|x} \}_{a=1}^{9}$. Each measurement operator $M_{a|x}$ can be expanded in the Gell-Mann basis as
\begin{equation}
M_{a|x} = \frac{1}{\sqrt{6}} \sum_{\mu=0}^{8} g_\mu^{\,a} \, G_\mu,
\end{equation}
where $G_0 = \sqrt{\frac{2}{3}} I$, and $\{G_\mu\}_{\mu=1}^{8}$ are the traceless Gell-Mann matrices satisfying $\mathrm{tr}(G_\mu G_\nu) = 2\delta_{\mu\nu}$.
Due to the trace constraint, each measurement operator can be uniquely represented by an 8-dimensional real vector
\begin{equation}
\vec g^{\,a} = (g_1^{\,a}, \dots, g_8^{\,a}) \in \mathbb{R}^8,
\end{equation}
which corresponds to the generalized Bloch representation for qutrit systems. A complete POVM is therefore represented as
\begin{equation}
\mathcal G_x = (\vec g^{\,1}, \vec g^{\,2}, \dots, \vec g^{\,9}) \in \mathbb{R}^{9 \times 8}.
\end{equation}

On this POVM configuration space, we construct a set of polynomial feature functions to describe the measurement dependence of the response function. For truncation order $D=2$, the feature family includes first- and second-order terms:
\begin{itemize}
\item First-order terms:
\begin{equation}
\{ g_i^{\,a} \}, \quad a=1,\dots,9,\; i=1,\dots,8,
\end{equation}

\item Second-order terms:
\begin{equation}
\{ g_i^{\,a} g_j^{\,b} \}, \quad a,b=1,\dots,9,\; i,j=1,\dots,8.
\end{equation}
\end{itemize}

These polynomial functions form a non-orthogonal feature family over the POVM configuration space. However, due to the high dimensionality and physical constraints, constructing an analytical orthonormal basis is generally intractable. To address this, we construct orthonormal basis functions using a numerically induced orthogonalization procedure. Specifically, we consider a set of POVM configurations $\{\mathcal G_x^{(k)}\}$ that are sampled from the physically admissible measurement space, satisfying positivity and completeness constraints. These samples are used to represent the underlying function space. We then define an empirical inner product on this sampled space:
\begin{equation}
\langle u, v \rangle = \frac{1}{M_{sample}} \sum_{k} u(\mathcal G_x^{(k)})\, v(\mathcal G_x^{(k)}),
\end{equation}
where $M_{sample}$ is the number of sampled configurations. Under this inner product, the polynomial feature family is orthogonalized using the Gram--Schmidt procedure, yielding an orthonormal basis set $\{B_m^2(\mathcal G_x)\}_{m=1}^{N}$. Finally, the response function is expressed as
\begin{equation}
f_a(\mathcal G_x,\lambda) = \sum_m c_a^m(\lambda)\, B_m^2(\mathcal G_x),
\end{equation}
which provides a tractable orthonormal expansion over the constrained high-dimensional POVM configuration space. This construction generalizes the spherical harmonic expansion in the qubit case to a polynomial basis expansion in the qutrit measurement space.

\section{Optimization of the LHS Model via Physics-Informed Machine Learning Method}

In the previous section, we established the physical parameterization of the LHS model through the hidden-variable matrix $\lambda$ and the hidden state $\sigma(\lambda)$. To determine whether a given entangled state $\rho$ is steerable, we now formulate the search for an optimal LHS model as a PIML task. The goal is to optimize the sets of hidden variables $\{\lambda_1, \lambda_2, \dots, \lambda_{N_{\text{hidden}}}\}$ and hidden states $\{\sigma(\lambda_1), \sigma(\lambda_2), \dots, \sigma(\lambda_{N_{\text{hidden}}})\}$ so that the model can reproduce the quantum assemblage $\{\sigma^{\text{QM}}_{a|x}\}_{a=1}^{O}$ across different measurement settings $M_x$. If such a model is successfully found, the state is certified as unsteerable. To this end, we introduce a loss function to quantify the discrepancy between the LHS and quantum assemblages. Subsequently, we demonstrate how this PIML method utilizes a JAX-powered gradient-based optimization loop to efficiently converge to the optimal LHS model.

\subsection{Loss Function}

To quantify the difference between the LHS assemblages $\{\sigma^{LHS}_{a|x}\}$ and the quantum assemblages $\{\sigma^{QM}_{a|x}\}$ for a fixed measurement $M_x$, we employ the trace distance as our metric. The trace distance between two density operators $\rho$ and $\rho'$ is defined as $D_Q(\rho,\rho')=\frac{1}{2}\lVert\rho-\rho'\rVert$, where $\lVert X \rVert:= \text{tr}[\sqrt{X^\dagger X}]$ is the trace norm \cite{PhysRevA.97.022338}. Accordingly, the distance between the two assemblages is:
\begin{align}
    D\left[\{\sigma^{LHS}_{a|x}\}_{a=1}^{O}, \{\sigma^{QM}_{a|x}\}_{a=1}^{O}\right] = \sum_{a=1}^O D_Q\left(\sigma^{LHS}_{a|x}, \sigma^{QM}_{a|x}\right).
\end{align}

Drawing upon machine learning terminology, our objective is to minimize a scalar loss function $\mathcal{L}$, where, as in Ref.~\cite{PRXQuantum.6.020317}, the deviation is averaged over the entire continuous measurement space:
\begin{align}
    \mathcal{L}(\text{LHS}||\text{QM}) = \left\langle D\left[\{\sigma^{LHS}_{a|x}\}_{a=1}^{O}, \{\sigma^{QM}_{a|x}\}_{a=1}^{O}\right] \right\rangle_{M_x}.
\end{align}

In this way, we treat all measurement choices as equally important. A converged loss function $\mathcal{L} = 0$ indicates that the quantum assemblages obtained under all possible measurements can be reproduced by the LHS model, thereby certifying the state $\rho$ as unsteerable at the computational level.

\subsection{Iterative Procedure for Solving the Optimal LHS Model}

The primary challenge in evaluating $\mathcal{L}$ lies in the continuous and infinite nature of the measurement space. To address this, we implement a dynamic ``sampling–evaluation–update'' iterative loop by adapting the procedure used in Ref.~\cite{PRXQuantum.6.020317}, in which a finite batch of measurements is sampled at each iteration to evaluate the loss and its gradient. By leveraging JAX's \texttt{vmap} technique for vectorized sampling and automatic differentiation for gradient computation, the optimization effectively explores the entire measurement space. The complete procedure is as follows:

(i) \textbf{Parameter Initialization}\par
The procedure requires the entangled state $\rho$, the system dimensions (qudit number $N$ and dimension $d$), and model precision (basis order $D$). Training configurations include the number of hidden variables $N_{\text{hidden}}$, total steps $N_{\text{steps}}$, measurements per step $N_{\text{meas/step}}$, and the learning rate $\eta$.

(ii) \textbf{Model Preparation and Vectorized Sampling} \par
This step prepares the quantum assemblages and response functions. 
To generate the target quantum assemblages $\{\sigma_{a|x}^{\text{QM}}\}$, we utilize JAX's \texttt{vmap} technique to achieve high-throughput sampling: in each iteration, $N_{\text{meas/step}}$ new measurement directions are dynamically drawn from the continuous space, and their corresponding unnormalized post-measurement states are computed efficiently in parallel.
Concurrently, the measurement response function $p(a|x, \lambda)$ is parameterized using the $N_{\text{basis\_dim}}$-dimensional orthonormal basis functions $\vec{B}^D(\vec{g})$, where $N_{\text{basis\_dim}}$ is determined by $D$ and represents the number of terms in the orthonormal basis. 
These basis functions are constructed from the Gell-Mann coefficient vectors $\vec{g}_a$ of the sampled measurement operators, integrated via the softmax activation: $p(a|x, \lambda) = \text{softmax}[f_a(x, \lambda)]$.

(iii) \textbf{LHS Model Assembly} \par
The LHS model is constructed using two primary sets of parameters: the hidden-variable matrix $\lambda$ with shape $(N_{\text{hidden}}, N_{\text{basis\_dim}})$ and the hidden states $\sigma(\lambda)$ with shape $(N_{\text{hidden}}, \sigma_{\text{param\_dim}})$, where $\sigma_{\text{param\_dim}} = d_{\text{Bob}}^2$ and $d_{\text{Bob}} = d^{N-1}$. As detailed in the parameterization section, each $\sigma(\lambda)$ is constructed via a Cholesky-like decomposition to strictly enforce positivity and unit-trace requirements by design. The LHS assemblage is then formed by averaging the state contributions over all hidden variables:
\begin{equation}
\sigma^{\text{LHS}}_{a|x} = \frac{1}{N_{\text{hidden}}} \sum_{i=1}^{N_{\text{hidden}}} p(a|x, \lambda_i) \sigma(\lambda_i).
\end{equation}

(iv) \textbf{Gradient-Based Optimization via Automatic Differentiation} \par

We minimize the sampled loss function $\mathcal{L} = \frac{1}{N_{\mathrm{meas/step}}} \sum_{\mathrm{meas}} D(\sigma^{LHS}, \sigma^{QM})$ through gradient descent. Leveraging JAX's automatic differentiation, we compute exact gradients $\nabla_{\boldsymbol{\theta}} \mathcal{L}$ with respect to the physical parameters $\boldsymbol{\theta} = (\lambda, \sigma)$. The parameters are updated as:
\begin{equation}
\boldsymbol{\theta}_{\mathrm{new}} = \boldsymbol{\theta}_{\mathrm{old}} - \eta \nabla_{\boldsymbol{\theta}} \mathcal{L}(\boldsymbol{\theta}_{\mathrm{old}}).
\end{equation}

where $\eta > 0$ is the learning rate. For a sufficiently small $\eta$, this update strictly follows the steepest descent direction on the physical parameter space, which reduces the loss as verified by a first-order expansion of $\mathcal{L}(\boldsymbol{\theta}_{\text{new}})$ with respect to $\eta$.

(v) \textbf{Result Output and Steerability Certification}\par
 This process cycles through $N_{\text{steps}}$ iterations output the optimized parameters $\lambda_{\text{opt}}$ and $\sigma_{\text{opt}}$, which define the final optimal LHS model. When the final loss value $\mathcal{L}$ converges close to zero, it signifies the successful construction of a valid LHS model that reproduces the target quantum assemblages across the continuous measurement space. This outcome thereby certifies the quantum state as unsteerable at the computational level.

\subsection{Parameter Scaling and Computational Complexity}

In this section, we analyze the parameter scaling and computational complexity of the proposed physically constrained LHS model within the Physics-Informed Machine Learning (PIML) framework, in order to assess its scalability with respect to the local dimension $d$. The total number of trainable parameters is determined by two main components: the parameterization of the response function and the representation of the local hidden quantum states. For the response function, each hidden variable $\lambda_i$ is associated with a set of coefficients $c_a^m(\lambda_i)$, which define an expansion over orthonormal basis functions $\{B_m\}_{m=1}^{N_{\text{basis\_dim}}}$ on the POVM configuration space. Here, $a=1,\dots,M$ denotes the measurement outcomes (with $M \le d^2$ for extremal POVMs), and $m=1,\dots,N_{\text{basis\_dim}}$ indexes the truncated orthonormal basis determined by the expansion order $D$. Therefore, each hidden variable corresponds to a parameter matrix of size $M \times N_{\text{basis\_dim}}$, leading to a total contribution of $N_{\text{hidden}} M N_{\text{basis\_dim}}$ parameters across all hidden variables. In addition, each hidden variable is associated with a local hidden quantum state $\sigma(\lambda_i)$ acting on Bob’s subsystem. This density matrix is parameterized via a Cholesky-type decomposition of a $d_B \times d_B$ matrix, where $d_B = d^{N-1}$ denotes the Hilbert space dimension of Bob’s subsystem, resulting in approximately $d_B^2$ real degrees of freedom per hidden variable. Combining both contributions, the total parameter complexity of the model scales as $\mathcal{O}\!\left(N_{\text{hidden}}(M N_{\text{basis\_dim}} + d_B^2)\right)$. Substituting $M = d^2$ and $d_B = d^{N-1}$ yields the explicit scaling with respect to the local dimension $d$, namely $\mathcal{O}\!\left(N_{\text{hidden}}(d^2 N_{\text{basis\_dim}} + d^{2(N-1)})\right)$.

The computational complexity is analyzed per training iteration, where one iteration corresponds to a single gradient update step over a batch of randomly sampled measurements. Therefore, the batch size $N_{\text{meas/step}}$ is already included in the per-step cost. The computational cost of the response function arises from evaluating the basis expansion
\[
f_a(\mathcal{G}_x,\lambda_i)=\sum_{m=1}^{N_{\text{basis\_dim}}} c_a^m(\lambda_i)\, B_m(\mathcal{G}_x)
\]
for all measurement samples, hidden variables, and measurement outcomes. This involves $N_{\text{meas/step}}$ measurement settings, $N_{\text{hidden}}$ hidden variables, $M$ outcomes, and $N_{\text{basis\_dim}}$ basis functions. In addition, the evaluation of the Gell-Mann polynomial features introduces a cost scaling approximately as $O(d^4)$ with respect to the local dimension. Therefore, the response function evaluation scales as
\[
O\!\left(N_{\text{meas/step}}\, N_{\text{hidden}}\, M\, N_{\text{basis\_dim}}\, d^4\right).
\]

The construction of hidden quantum states $\sigma(\lambda_i)=M_\lambda M_\lambda^\dagger$ requires matrix multiplication of $d_B \times d_B$ matrices, resulting in a computational cost of $O(d_B^3)$ per hidden variable, and thus a total cost of
\[
O\!\left(N_{\text{hidden}} d_B^3\right).
\]

Furthermore, the evaluation of the loss function involves computing trace distances between LHS-generated states and quantum assemblage states, which requires eigenvalue or singular value decompositions of $d_B \times d_B$ density matrices. This leads to a cost of
\[
O\!\left(N_{\text{meas/step}}\, M\, d_B^3\right).
\]

Combining all contributions, the total computational complexity per iteration can be written as
\begin{equation}
\begin{split}
\mathcal{C}_{\text{step}} = O \Big( & N_{\text{meas/step}} N_{\text{hidden}} M N_{\text{basis\_dim}} d^4 \\
& + N_{\text{hidden}} d_B^3 + N_{\text{meas/step}} M d_B^3 \Big),
\end{split}
\end{equation}
where $M = d^2$ and $d_B = d^{N-1}$. Consequently, the total training complexity over $N_{\text{steps}}$ iterations is given by
\[
\mathcal{C}_{\text{total}} = N_{\text{steps}} \cdot \mathcal{C}_{\text{step}}.
\]

This analysis shows that the dominant computational bottlenecks arise from the high-order polynomial feature evaluation in the measurement space (scaling as $d^4$) and the cubic matrix operations required for constructing and diagonalizing the hidden quantum states (scaling as $d_B^3$), with the latter typically dominating in higher-dimensional multipartite scenarios.

\section{The Applications}
To validate the effectiveness of our PIML method, we apply it to determining the steerability of two-qubit Werner states and two-qutrit isotropic states under different measurement scenarios.

\subsection{Two-qubit werner states}
We take the two-qubit Werner state as an example:
\begin{align}
&\rho(v) = v|\psi^-\rangle\langle\psi^-| + \frac{(1-v)}{4} I_2 \otimes I_2, \\ \notag
&|\psi^-\rangle = \frac{1}{\sqrt{2}}(|01\rangle - |10\rangle).
\end{align}

The parameter \(v \in [0, 1]\) represents the visibility of the singlet. It is known that these states are separable for  \(v \leq 1/3\)  \cite{PhysRevLett.77.1413}. As the parameter \(v\) decreases from 1 to 0, the quantum state gradually transitions from a maximally entangled state to a completely separable state, and its steerability correspondingly changes from ``steerable" to ``unsteerable".

For the entangled state, its steering properties may differ across different measurement scenarios. To this end, the following will examine the critical value of $v$ under three different measurement scenarios, including: three fixed Pauli measurements, arbitrary PVMs, and arbitrary POVMs.
\begin{figure}[htbp]
    \centering
     \includegraphics[width=\linewidth]{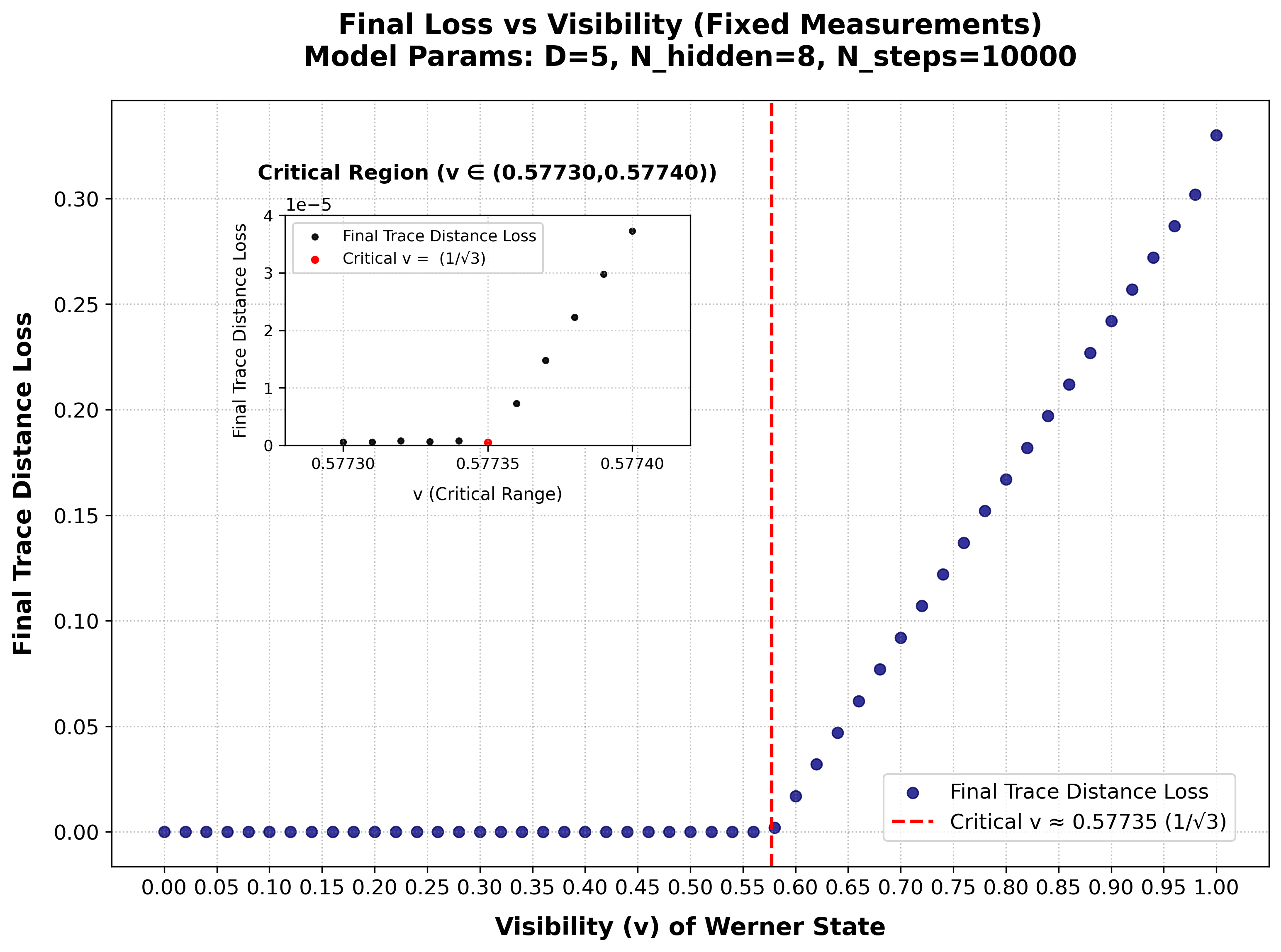}  

    \caption{Steerability of the two-qubit Werner states under three Pauli measurements. The parameters adopted for optimizing the LHS model are as follows: the order of orthonormal basis functions   \(D = 5\), the number of hidden variables \(N_{\text{hidden}} = 8\), and the number of gradient descent iterations   \(N_{\text{steps}} = 10^5\). Theoretically, the higher the order   D   of the orthonormal basis functions and the more gradient descent iterations, the higher the result accuracy. However, practical verification shows that the currently selected parameters are sufficient to meet the requirements.}
    \label{Result1}
\end{figure}

\subsubsection{The Three fixed Pauli Measurements}

Theoretical analysis shows that when Alice employs the three Pauli measurements $X$, $Y$, and $Z$, the corresponding steering inequality for this system is:
\[
\langle XX\rangle + \langle YY\rangle + \langle ZZ\rangle \leq \sqrt{3}.
\]

This inequality gives a clear steering threshold: the Werner state is unsteerable when $v \leq 1/\sqrt{3}$ \cite{PhysRevA.80.032112}.

We apply our method to analyze the steerability of this quantum state under three fixed measurements. We construct an LHS model for the assemblages on Bob's side. Here, Alice performs three fixed measurements, eliminating the need for random measurement sampling during training to obtain optimal parameters. The loss function only needs to average over these three measurements. Our numerical results accurately reproduce the theoretical critical point $v_c = \frac{1}{\sqrt{3}}$, as shown in the Fig.\ref{Result1}.

\subsubsection{The Arbitrary PVMs}
The LHS models constructed based on a finite set of measurements (e.g., three Pauli measurements) cannot provide a complete criterion for the steerability of quantum states. The threshold obtained under three Pauli measurements (\(v_c = 1/\sqrt{3}\)) is far from optimal. Next, we consider more PVMs, which are experimentally feasible to implement. Their measurement operators must satisfy hermiticity \(M_{a|x}^\dagger = M_{a|x}\), positive semi-definiteness \(M_{a|x} \geq 0\), idempotency \(M_{a|x}^2 = M_{a|x}\), and orthogonality \(M_{a|x} M_{a'|x} = \delta_{a,a'}M_{a|x}\).

\begin{figure}[htbp]
    \centering
    \includegraphics[width=\linewidth]{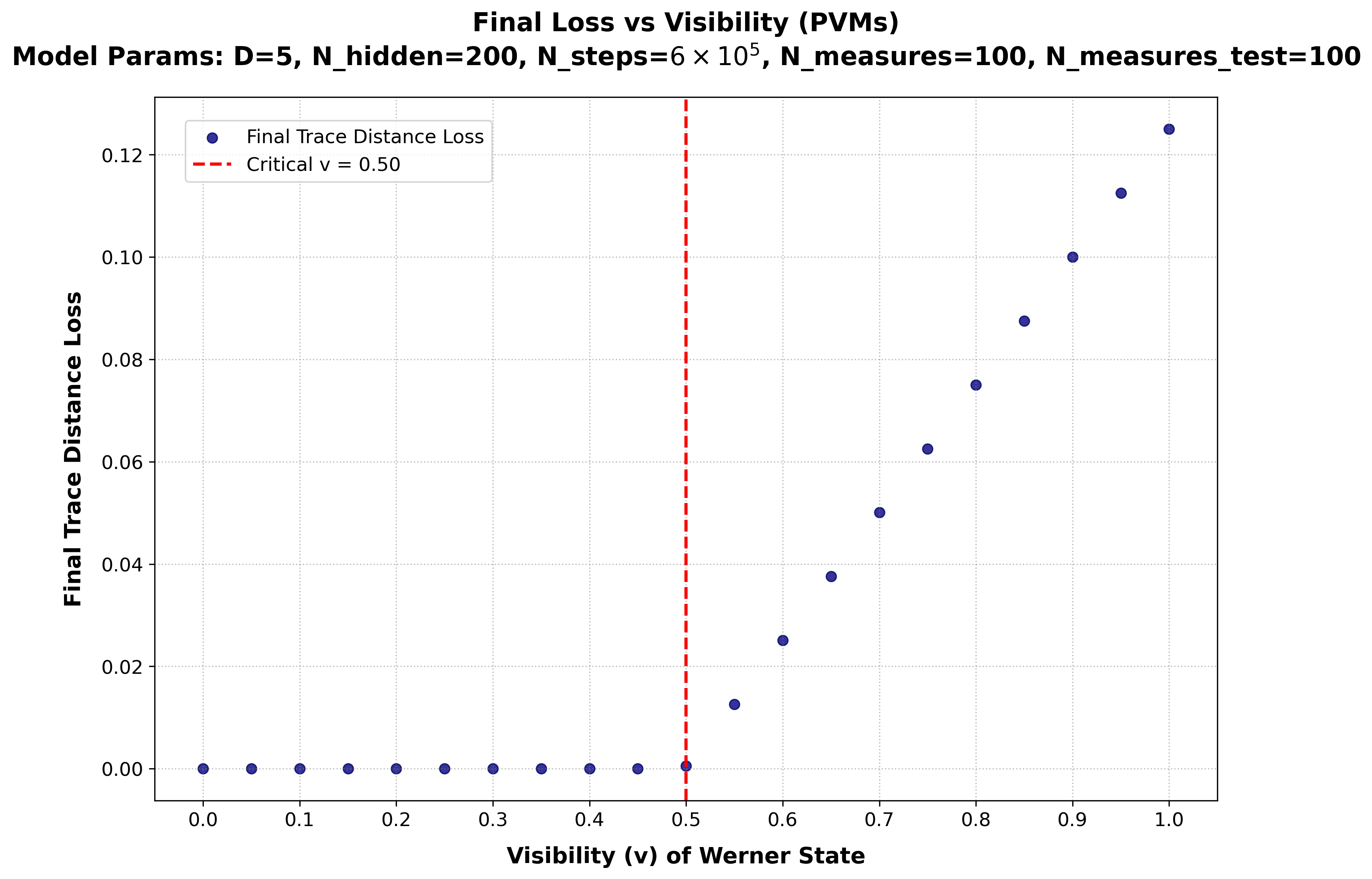}
    \caption{Steerability of the two-qubit Werner states under PVMs. The parameters adopted for optimizing the LHS model are as follows: the order of orthonormal basis functions $D = 5$, the number of hidden variables $N_{\text{hidden}} = 200$, and the number of gradient descent steps $N_{\text{steps}} = 6 \times 10^5$. To verify that the optimal LHS model is not only applicable to the measurements used in training, we uniformly and randomly sampled a batch of measurements for testing, which confirms the model's universality.}
    \label{Result2}
\end{figure}

We perform uniform random sampling over the entire space of qubit PVMs. At each step of the gradient descent, we resample a new batch of PVMs. Theoretically, when the number of optimization steps and sample size are sufficiently large, this strategy can effectively approximate coverage of the entire continuous measurement space. For the quantum assemblages generated by sampling, we simultaneously optimize the hidden variables \(\lambda\) and hidden states \(\sigma(\lambda)\) to find the optimal local hidden-state model. To rigorously verify the universality of the obtained model and ensure it is not merely tailored to the finite training measurement set, we compute the loss function on an independent test set. Our numerical results confirm the theoretical conclusion: for two-qubit Werner states, the assemblages generated under PVMs admit an LHS model up to \(v = \frac{1}{2}\) (see Fig. \ref{Result2}).
\subsubsection{The Arbitrary POVMs}
PVMs represent a special case of generalized POVMs. The existence of non-projective POVMs enables quantum theory to generate statistical predictions that cannot be simulated using only projective measurements and classical randomness \cite{DAriano_2005}. Therefore, restricting quantum steering research solely to PVMs is incomplete. We next investigate the steerability of entangled states under the more general POVMs, where the measurement operators \(M_{a|x}\) satisfy hermiticity \(M_{a|x}^\dagger = M_{a|x}\) and positive semi-definiteness \(M_{a|x} \geq 0\).

\begin{figure}[htbp]
    \centering
    \includegraphics[width=\linewidth]{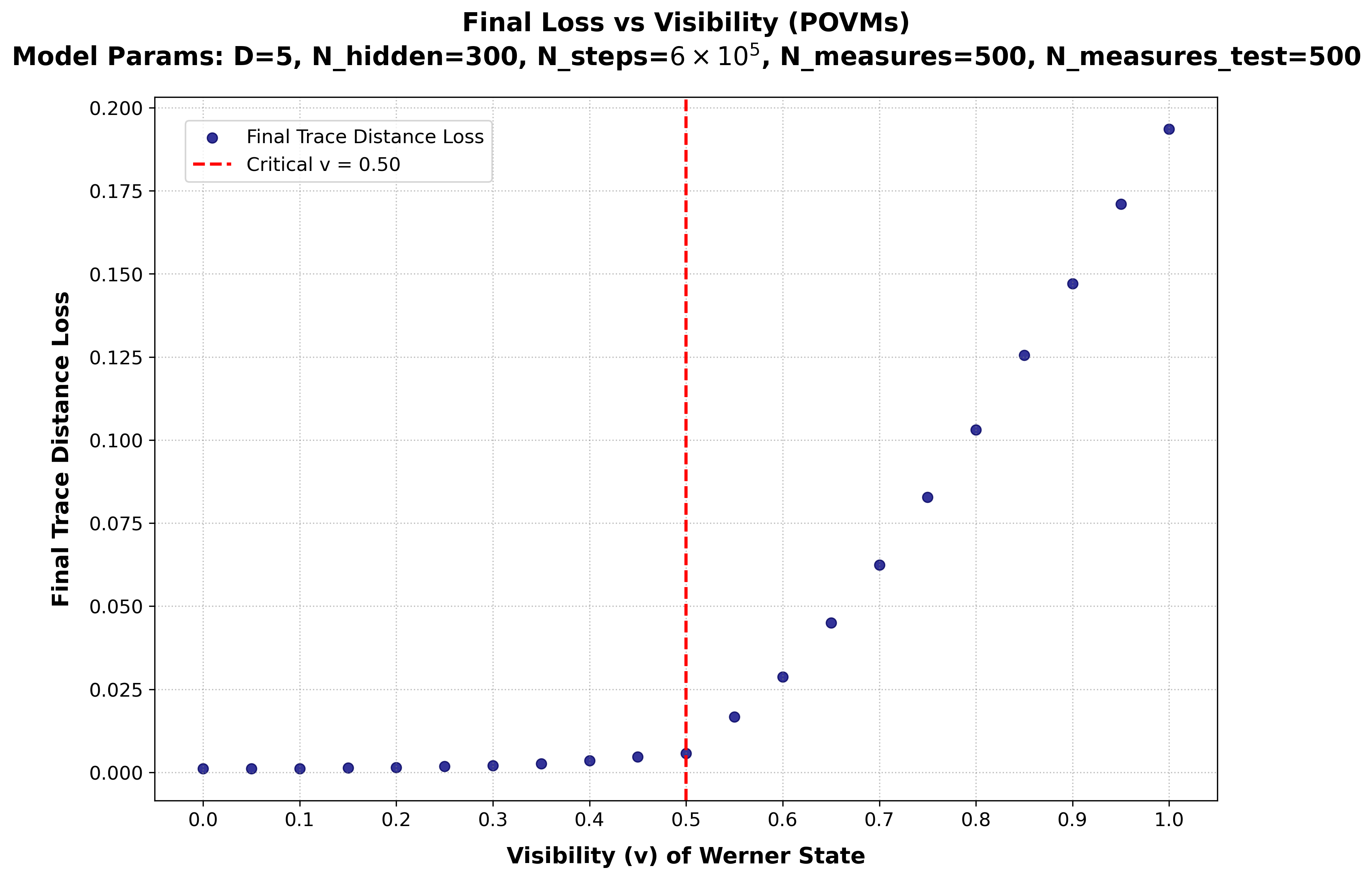}
    \caption{Steerability of the two-qubit Werner states under POVMs. The parameters adopted for optimizing the LHS model are as follows:  the order of orthonormal basis functions \(D = 2\), the number of hidden variables \(N_{\text{hidden}} = 300\), and the number of gradient descent steps \(N_{\text{steps}} = 10^5\). At each step of the gradient descent, \(N_{\text{measures}}\) measurements are randomly selected to optimize the parameters, yielding the optimal \(\lambda\) and \(\sigma(\lambda)\). Finally, the loss value is computed using \(N_{\text{measures\_test}}\) measurements.}
    \label{Result3}
\end{figure}

Our method is the first numerical approach to verify the critical value \(v=1/2\) for the steerability of two qubit werner states via the construction of an LHS model under POVMs (see Fig. \ref{Result3}). Under the configuration of Python 3.12, CUDA 12.4, and an RTX 4090 GPU, obtaining the optimal LHS model for the entangled state corresponding to each $v$ value requires one hour of computation per run.

\subsection{Two-qutrit isotropic states}

We investigate the steerability of bipartite qutrit states under PVMs and POVMs. Specifically, we consider isotropic states of the form:
\begin{align}
    &\rho = v|\psi^+\rangle\langle\psi^+| + \frac{(1-v)}{9} I_3 \otimes I_3, \\ \notag
    &|\psi^+\rangle = \frac{1}{\sqrt{3}}(|00\rangle + |11\rangle + |22\rangle).
\end{align}
Here $|\psi^+\rangle$ denotes the maximally entangled isotropic state. It is well known that these states are entangled for $v > \frac{1}{4}$, and are steerable under PVMs for $v > \frac{5}{12}$ \cite{PhysRevLett.98.140402}.

\begin{figure}[htbp]
    \centering
    \includegraphics[width=\linewidth]{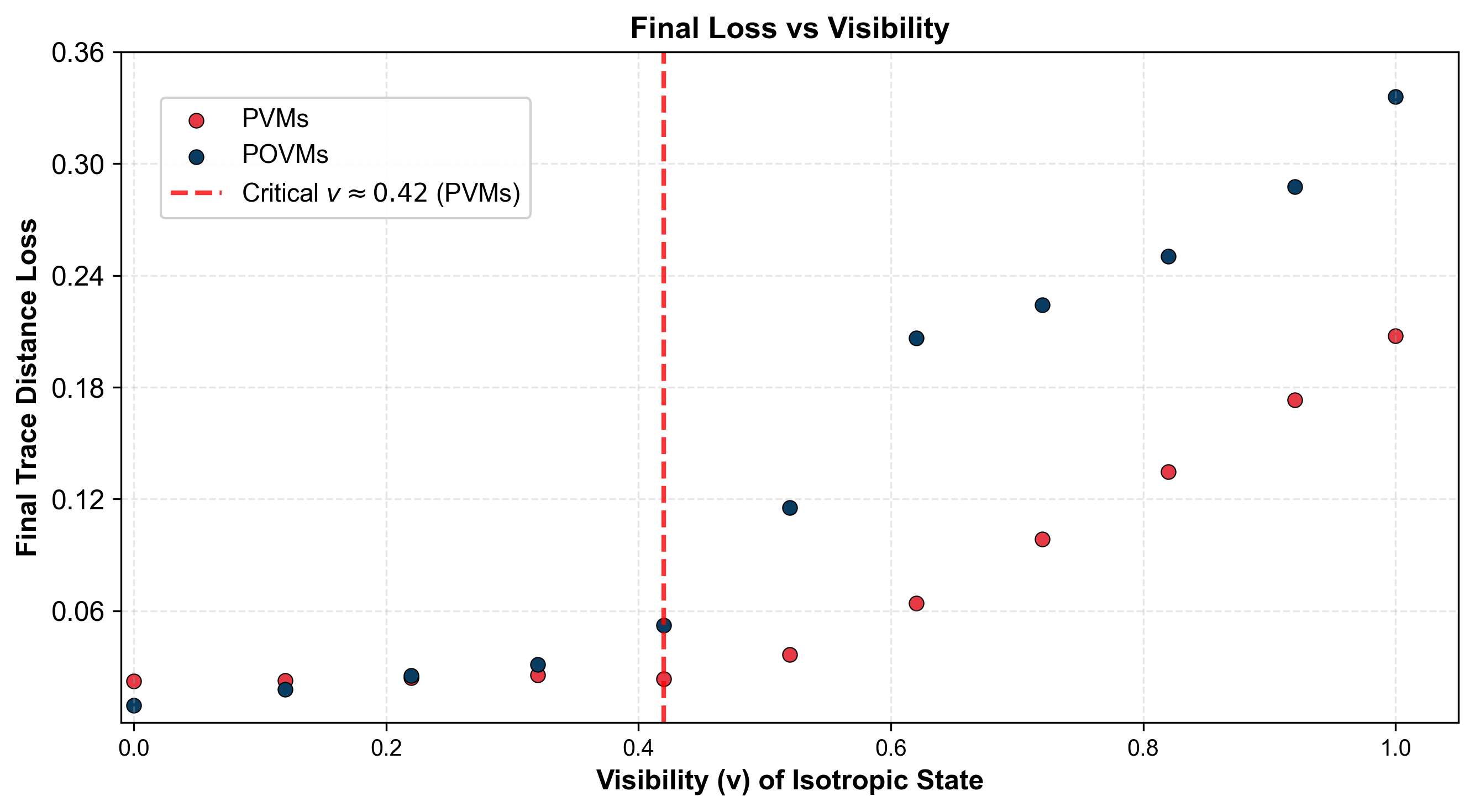}
    \caption{Steerability of the two-qutrit isotropic states under PVMs and POVMs. Red points show results under PVMs, with the following parameters: order of orthonormal basis functions \(D = 2\), number of hidden variables \(N_{\text{hidden}} = 100\), number of gradient descent steps \(N_{\text{steps}} = 2 \times 10^5\), and \(N_{\text{meas}} = 200\) randomly selected measurements per gradient step. Blue points correspond to results under general POVMs, with parameters: \(D = 3\), \(N_{\text{hidden}} = 200\), \(N_{\text{steps}} = 2 \times 10^5\), and \(N_{\text{meas}} = 300\).}
    \label{Result4}
\end{figure}

For the case of PVMs, our numerical results indicate that the steering transition occurs at a critical value of \( v = 0.42 \), which is in very close agreement with the known theoretical bound \( v > \frac{5}{12} \). For general POVMs, we observe that the growth rate of the loss function increases significantly for \( v > 0.3 \), with a further acceleration occurring beyond \( v > 0.4 \) (see Fig.~\ref{Result4}). Based on this observed trend, we conjecture that the steering threshold under general POVMs is close to \( v \approx 0.3 \). This indicates that, compared with PVMs, POVMs possess a certain advantage in detecting the steerability of such high-dimensional quantum states.

\begin{figure}[htbp]
    \centering
    \begin{subfigure}[b]{\linewidth}
        \centering
        \includegraphics[width=\linewidth]{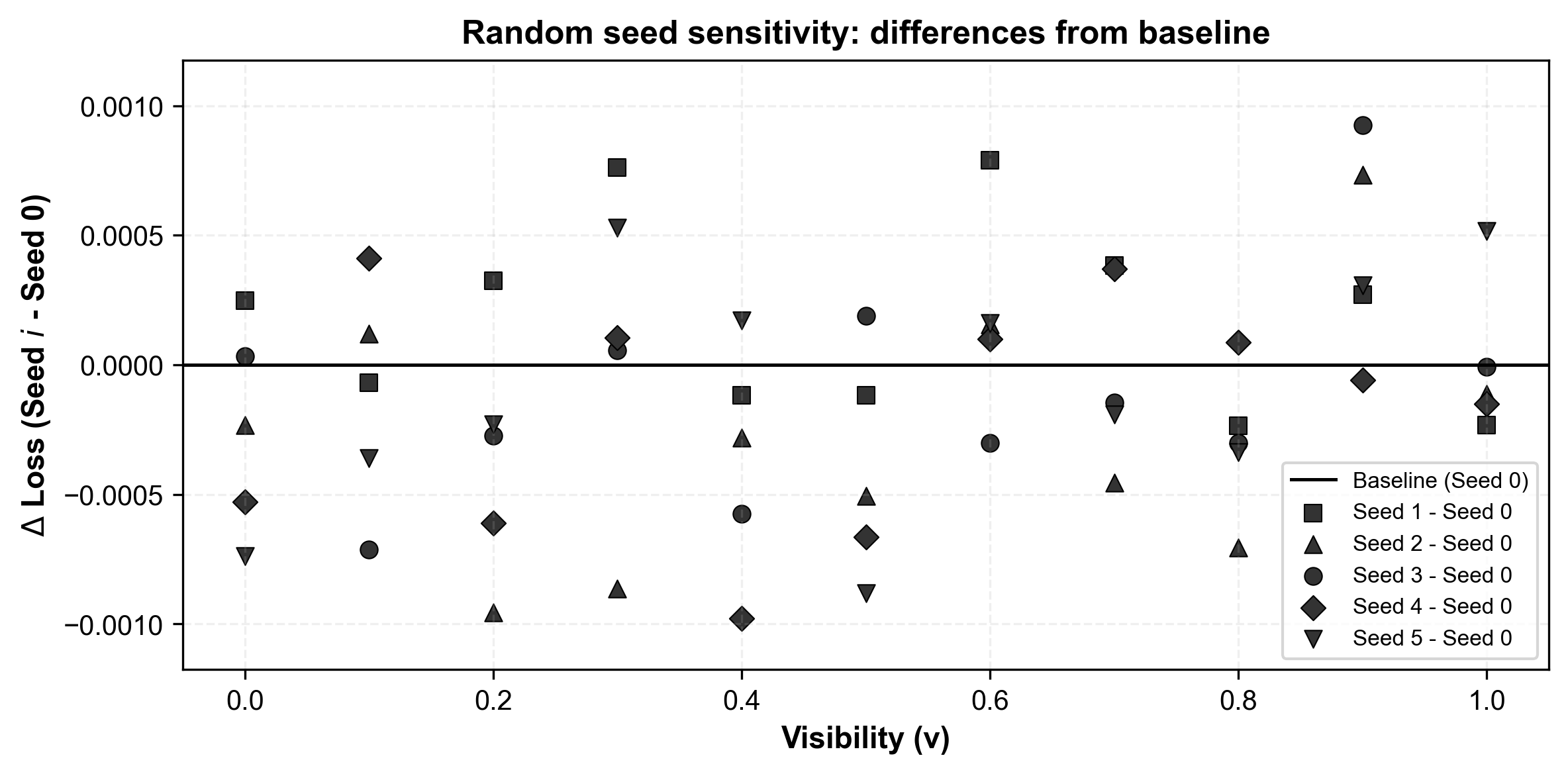}
        \caption{Difference in loss across different random seeds.}
        \label{fig:seed_sensitivity}
    \end{subfigure}
    \vspace{0.5em}
    
    \begin{subfigure}[b]{\linewidth}
        \centering
        \includegraphics[width=\linewidth]{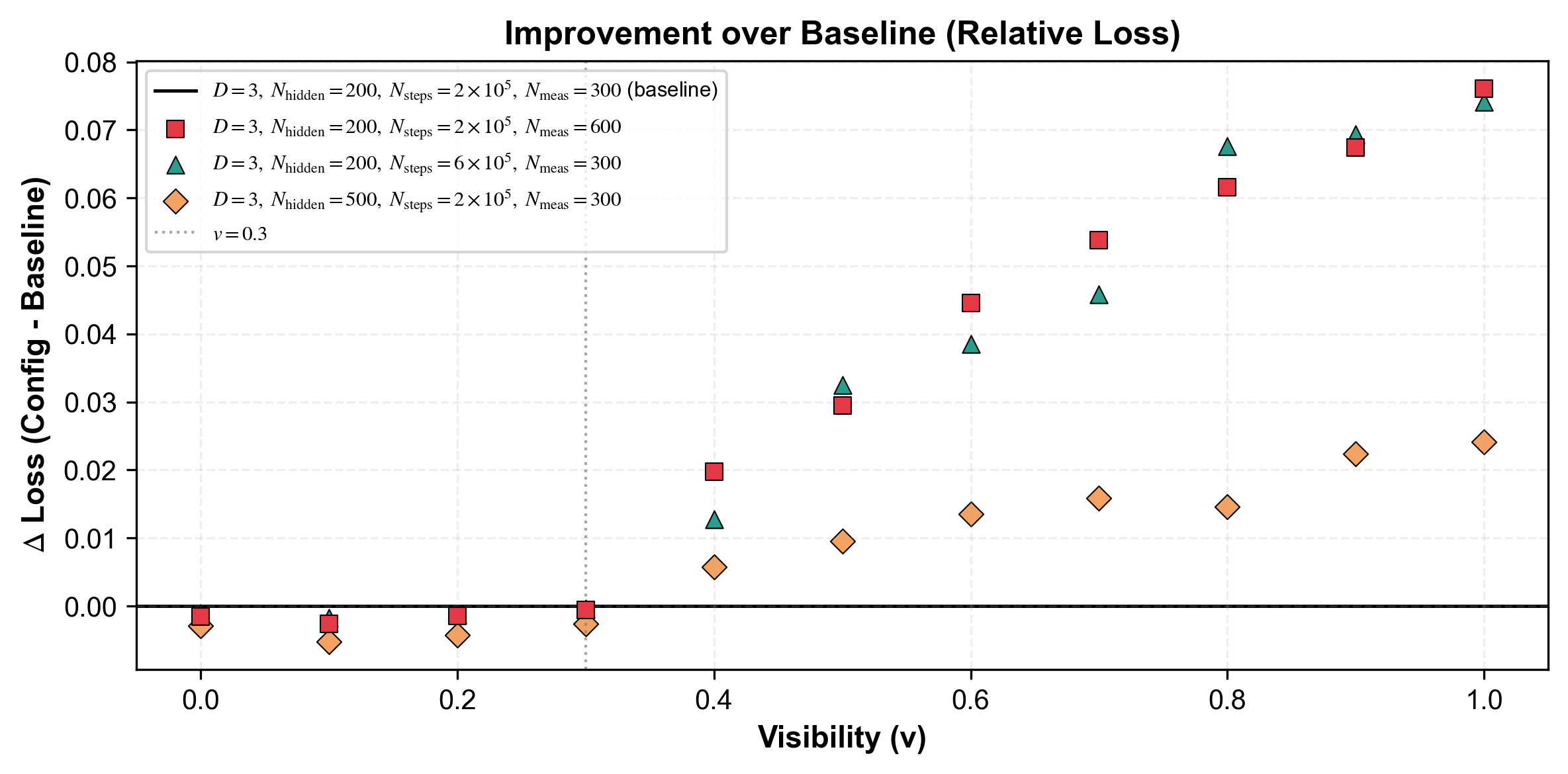}
        \caption{Difference in loss when increasing $N_{\text{meas}}$, $N_{\text{steps}}$, or $N_{\text{hidden}}$ relative to the $D=3$ baseline.}
        \label{fig:hyperparam_diff}
    \end{subfigure}
    \vspace{0.5em}
    
    \begin{subfigure}[b]{\linewidth}
        \centering
        \includegraphics[width=\linewidth]{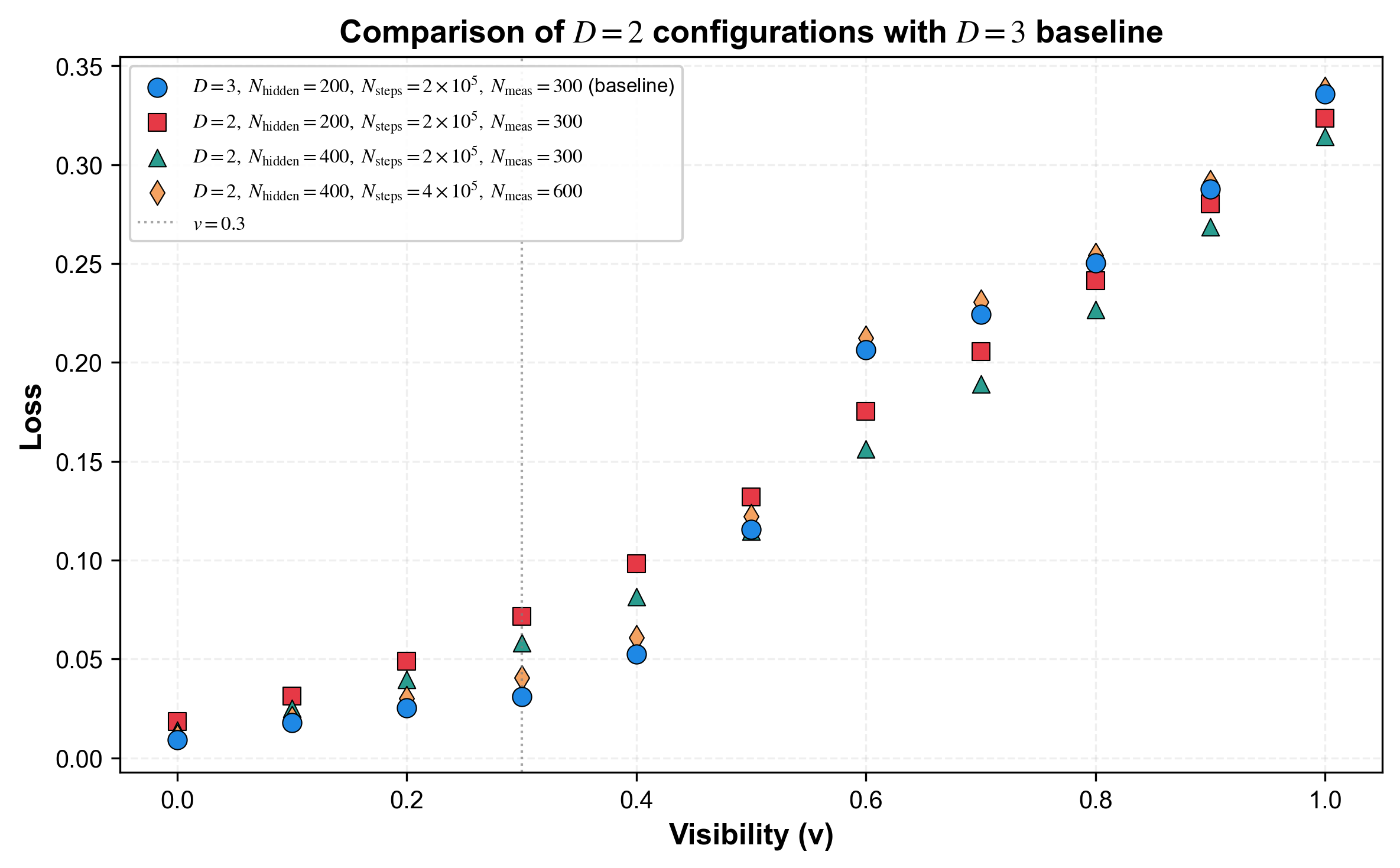}
        \caption{Absolute loss for $D=2$ configurations compared to the $D=3$ baseline.}
        \label{fig:D2_comparison}
    \end{subfigure}
    \caption{Numerical robustness tests. (a) Difference in loss across six different random seeds; (b) difference in loss when varying $N_{\text{hidden}}$, $N_{\text{steps}}$, $N_{\text{meas}}$ under $D=3$; (c) comparison of basis truncation orders $D=2$ and $D=3$.}
    \label{fig:robustness}
\end{figure}

Since the identification of the local hidden-state model is inherently a non-convex optimization problem, there is a potential risk of converging to a suboptimal local minimum, which could lead to unreliable simulation bounds. To rule out such numerical artifacts, we systematically evaluated the sensitivity of our results to several key numerical configurations, including the random seed, hyperparameters ($N_{\text{hidden}}$, $N_{\text{steps}}$, $N_{\text{meas}}$), and the basis truncation order $D$. The results are summarized in Fig.~\ref{fig:robustness}.

As shown in Fig.~\ref{fig:seed_sensitivity}, we repeated the optimization with six different random seeds while keeping all other parameters fixed ($D=3$, $N_{\text{hidden}}=200$, $N_{\text{steps}}=2\times10^5$, $N_{\text{meas}}=300$). The loss differences relative to the first seed are all within $\pm0.001$, i.e., two orders of magnitude smaller than the absolute loss values. This indicates that the optimization is essentially insensitive to the initial seed and that different seeds converge to nearly the same loss landscape. Fig.~\ref{fig:hyperparam_diff} examines the influence of increasing $N_{\text{meas}}$ to 600, $N_{\text{steps}}$ to $6\times10^5$, or $N_{\text{hidden}}$ to 500 (while keeping $D=3$ and the other parameters at their baseline values). The relative differences are at most $0.03$, and the characteristic steepening of the loss curve near $v=0.3$ persists in all cases. Therefore, our conclusion about the transition point is robust against reasonable variations of these hyperparameters. Fig.~\ref{fig:D2_comparison} compares the loss obtained with $D=2$ (under several increased resource settings) against the $D=3$ baseline. For $D=2$, the loss values are systematically higher and do not exhibit a clear change of slope around $v=0.3$; even after substantially increasing $N_{\text{hidden}}$, $N_{\text{steps}}$, and $N_{\text{meas}}$, the $D=2$ results remain far from the $D=3$ baseline and lack a distinct transition. This demonstrates that a truncation order of $D=3$ is necessary to capture the genuine steering behavior, and that the observed $v\approx0.3$ transition is not an artifact of the basis truncation.

In summary, although an exhaustive search over all possible measurement combinations is beyond our current resources, the three independent numerical checks — random seeds, hyperparameter variations, and basis truncation order — all consistently show that the transition near $v\approx0.3$ is robust. Therefore, this result reflects the intrinsic steerability of the two-qutrit isotropic states under general POVMs, rather than a numerical artifact. Finally, it should be noted that the numerical precision achievable with our approach is limited by computational resources. This limitation mainly arises because, in high-dimensional systems, an increase in the number of measurement outcomes requires the introduction of a larger set of orthonormal basis functions for their description, which significantly enlarges the matrix size associated with each hidden variable $\lambda$, thereby making the numerical optimization considerably more challenging. In our computational environment (Python 3.12, CUDA 12.4, NVIDIA RTX 4090 GPU), a single optimization run to obtain the optimal local hidden-state model for a given visibility $v$ under the $D=3$ parameter setting typically requires several tens of hours of computation.

\section{conclusion}
In this work, we propose a general physics-informed machine learning method for constructing LHS models to determine the steerability of arbitrary entangled states. The central challenge lies in exploring all possible measurement settings in the continuous measurement space, which is required to certify steerability in full generality. We address this challenge by combining real-time sampling enabled by JAX’s vmap with an orthonormal basis expansion of the response functions across measurement settings, building on the approach introduced for constructing LHV models in Ref.~\cite{PRXQuantum.6.020317}. Furthermore, we introduce an unconstrained complex matrix  $M_\lambda$, which is mapped to a valid hidden state $\sigma(\lambda)$  through parameter reparameterization, thereby enforcing physical consistency directly within the model architecture. Within this framework, the construction of an LHS model is naturally reformulated as a differentiable parameter optimization problem over batches of high-throughput measurements. By leveraging automatic differentiation to evaluate gradients for gradient-based optimization, the model parameters are iteratively updated until convergence to an optimal LHS model. When the loss function reaches zero, the resulting model exactly reproduces the target quantum assemblages, certifying the state as unsteerable at the computational level.

The strength of our approach lies in its explicit embedding of physical constraints within the learning architecture, enabling it to overcome the limitations of previous numerical methods and reliably reproduce analytical steering bounds. For certain noisy entangled states, existing numerical approaches are often limited by their measurement search strategies or model expressiveness. As a result, they cannot fully identify measurement settings that rule out LHS descriptions and are limited in the range of states for which steerability can be assessed \cite{PhysRevLett.117.190401, PhysRevLett.117.190402, PhysRevA.111.052446}. In contrast, the PIML method proposed here remains capable of correctly identifying steerability in these challenging regimes. For two-qubit Werner states under three Pauli measurements \cite{PhysRevA.80.032112}, arbitrary PVMs \cite{PhysRevA.93.022121}, and arbitrary POVMs \cite{PhysRevLett.132.250201}, our method accurately reproduces all known analytical visibility bounds. For two-qutrit isotropic states, we likewise recover the analytical steering visibility under arbitrary PVMs \cite{PhysRevLett.98.140402}. Moreover, in the absence of analytical results for the steerability of isotropic states under POVMs, our method provides the first numerical insights into this regime. Due to computational limitations, the critical visibility can only be roughly estimated as \( v \approx 0.3 \), providing further evidence that POVMs may offer an advantage over PVMs in revealing the steerability of these states.

The accuracy of our general PIML method can, in principle, be further improved by increasing key parameters such as the polynomial expansion order \(D\), the number of hidden variables \(N_{\mathrm{hidden}}\), and the number of gradient descent steps \(N_{\mathrm{steps}}\). However, such improvements in accuracy are typically accompanied by a significant increase in computational cost. With the future development of more efficient optimization algorithms and the continuous growth of computational resources, our method can be further adapted to achieve more effective and precise certification of steerability.

\section*{Acknowledgment}
This work is supported by the National Natural Science Foundation
of China (Grants No. 62571060, No. 62171056, and No. 62220106012).

\bibliography{References}

\end{document}